\documentclass{aastex}

\newcommand{\kms}{${\rm km \: s^{-1}}$}
\newcommand{\lsim}{${<}\atop{\sim}$}
\newcommand{\gsim}{${>}\atop{\sim}$}
\newcommand{\SI}{S~{\sc i}\ }
\newcommand{\FeI}{Fe~{\sc i}\ }
\newcommand{\FeII}{Fe~{\sc ii}\ } 
\newcommand{\Teff}{$T_{\rm eff}$\ }
\newcommand{\logg}{log $g$\ }
\newcommand{\Wlam}{$W_{\lambda}$\ }
\newcommand{\Francois}{Fran{\c c}ois\ }

\slugcomment{Submitted to Astrophys. J.}

\shorttitle{Sulfur in Metal-Poor Stars}
\shortauthors{Takada-Hidai et al.}

\begin{document}

\def\lsim{${<}\atop{\sim}$}
\def\gsim{${>}\atop{\sim}$}


\title{Behavior of Sulfur Abundances in Metal-Poor Giants and Dwarfs}


\author{Masahide Takada-Hidai}
\affil{Liberal Arts Education Center, Tokai University,
1117 Kitakaname, Hiratsuka, Kanagawa, Japan 259-1292}
\email{hidai@apus.rh.u-tokai.ac.jp}

\author{Yoichi Takeda}
\affil{Institute of Astronomy, The University of Tokyo,
    Mitaka, Tokyo, Japan 181-0015}
\email{takedayi@cc.nao.ac.jp}

\author{Shizuka Sato}
\affil{Department of Aeronautics, School of Engineering, Tokai University,
1117 Kitakaname, Hiratsuka, Kanagawa, Japan 259-1292}
\email{shizuka@apus.rh.u-tokai.ac.jp}

\author{Satoshi Honda}
\affil{National Astronomical Observatory, Mitaka, Tokyo, Japan
 181-8588}
\email{honda@optik.mtk.nao.ac.jp}

\author{Kozo Sadakane}
\affil{Astronomical Institute, Osaka Kyoiku University, Kashiwara, Osaka,
Japan 582-8582}
\email{sadakane@cc.osaka-kyoiku.ac.jp}

\author{Satoshi Kawanomoto}
\affil{Department of Astronomy, School of Science, The University of Tokyo,
Bunkyo-ku, Tokyo, Japan 113-0033}
\email{kawanomo@optik.mtk.nao.ac.jp}

\author{Wallace L. W. Sargent and Limin Lu}
\affil{Department of Astronomy, California Institute of Technology,
Mail Stop 105-24, Pasadena, CA 91125}
\email{wws@astro.caltech.edu}

\and 

\author{Thomas A. Barlow}
\affil{Infrared Processing and Analysis Center, California Institute of 
Technology, Mail Stop 100-22,
Pasadena, CA 91125}
\email{tab@ipac.caltech.edu}


\begin{abstract}
LTE and NLTE abundances of sulfur in 6 metal-poor giants and 61 dwarfs 
(62 dwarfs, including the Sun)
were explored in the range of $-3$ \lsim [Fe/H] \lsim $+0.5$
using high-resolution, high signal-to-noise ratio spectra of the
\SI 8693.9 \AA\  and 8694.6 \AA\  lines observed by us and
measured by \Francois (1987, 1988) and Clegg et al. (1981).
NLTE effects in S abundances
are found to be small and practically negligible.
The behavior of [S/Fe] vs. [Fe/H] exhibits a linear increasing trend
without plateau with decreasing [Fe/H].  Combining our results with
those available in  the literature, we find that the slope of the increasing
trend is  $-0.25$ in the NLTE behavior of [S/Fe], which is
comparable to that observed in [O/Fe].
The observed behavior of S  may require chemical
evolution models of the Galaxy, in which scenarios of hypernovae 
nucleosynthesis and/or time-delayed deposition into interstellar medium
are incorporated.

\end{abstract}




\section{Introduction} 

The behavior of $\alpha$-elements such as oxygen, magnesium, silicon, 
sulfur, calcium, and titanium in metal-poor halo and disk stars provides
 us with 
very useful information and constraints for exploring 
stellar nucleosynthesis and chemical evolution of the Galaxy in the early
phase, since these elements are believed to be produced mainly by type II
supernovae (SNe II) of massive stars (e.g. Timmes, Woosely, \& Weaver 1995;
Chiappini et al. 1999; Carretta, Gratton, \& Sneden 2000) and are fossilized
in stars which have been born at various epochs of the Galaxy evolution.

Numerous abundance works have been carried out for these elements (e.g.
 see a review of McWilliam 1997). It is well known that the 
$\alpha$-elements, except for O and S, show a general trend of
abundances such that  [$\alpha$-element/Fe] increases with decreasing 
metallicity [Fe/H] down to
about $-1$ dex, and forms a plateau with  [$\alpha$-element/Fe] $\sim 0.3$ --
0.5 dex in the range of [Fe/H] $< -1$ dex.
Here we define [A/B]$\equiv$  log (A/B)$_{\rm star} - $log (A/B)$_{\rm Sun}$ 
for the elements of A and B and use it throughout the text hereafter. 
The behavior of [O/Fe] against [Fe/H] is controversial, i.e.  whether 
[O/Fe] shows a plateau (e.g. Barbuy 1988; Fulbright \& Kraft 1999) or 
a steady linear increase with decreasing [Fe/H] (e.g. Boesgaard et al. 1999;
Israelian et al. 2001; Takeda et al. 2002). 
     
As for the behavior of S  among metal-poor stars,
it is not well understood because there are only four
previous works dealing with a larger number of stars:
Clegg,  Lambert, \& Tomkin (1981) and 
Fran{\c c}ois (1987, 1988) for a total number of 44 dwarfs 
 and one giant star; Prochaska et al. (2000) for nine thick 
disk dwarfs.
  These works analyzed  S
abundances in the metallicity range, $-1.6$ \lsim [Fe/H] \lsim $+0.5$,
but mostly in [Fe/H] $> -1$.
The scarcity of the S abundance works is mainly  due to the difficulty in
observing the S lines available to abundance analysis. 
There are no strong lines
in the visual region suitable enough for analysis, but there are a few 
relatively strong lines
of \SI  (multiplet number 6) in the near-infrared region of 8670 -- 8700 \AA.
Among the \SI(6) lines, the strongest line at 8694.641 \AA\  and a weak line
at  8693.958 \AA\  are 
free from blending with lines of other ions such as Si {\sc i} and \FeI,
while the remaining lines are blended with 
such ions. Even the strongest  line at 8694.6 \AA\  generally becomes very 
weak for stars with effective temperature (\Teff)  of 
$T_{\rm eff} < $  5500 K 
 and with metallicity of [Fe/H] $< -0.5$ (cf. Fran{\c c}ois 1987, 1988).

Among these previous studies, \Francois (1988) suggested that [S/Fe] becomes
overabundant and constant in the halo stars with a value of $+0.6$ dex in 
the metallicity range of $-1.6 <$[Fe/H] \lsim $-1$, 
while Prochaska et al. (2000) found no indication of a significant 
overabundance of [S/Fe] with a mean value of $+0.11\pm0.08$ dex
 and no trend with metallicity in the 
range of $-1.2$ \lsim [Fe/H] \lsim $-0.3$. Prochaska et al. (2000)
also postulated that a more careful, extensive stellar abundance analysis
of S in metal-poor stars is warranted since S is one of important elements
in quasar absorption-line studies.

Very recently, Israelian \& Rebolo (2001) analyzed the S abundances in
six metal-poor stars with metallicities in the range of $-3 <$ [Fe/H] $<-0.2$
using observational data of \SI lines at 8694 \AA. Their results indicate
a monotonic increase of [S/Fe] as [Fe/H] decreases, reaching [S/Fe]$ \sim 0.7$
 -- $0.8$ below [Fe/H] $= -2$.

Takeda et al. (2002; hereafter Paper I)  also very recently reported the 
abundance  analysis of S for very metal-poor giants with [Fe/H] \lsim $-1.5$
observed  with the High Resolution Echelle Spectrometer (HIRES; Vogt 1994) of the 
10 m Keck I telescope as a byproduct in the oxygen 
abundance study.
 It is suggested that [S/Fe] increases linearly  with decreasing [Fe/H],
which is just the same trend as found in Israelian \& Rebolo (2001),
and this behavior resembles that of O. 
However, Paper I's results of S are yet preliminary and  should be
superseded with those of this study. 

In this paper, in order to explore the behavior of S in the metallicity range
 of $-3$ \lsim [Fe/H] \lsim $+0.5$, we carried out extensive 
LTE abundance analyses of S for the six samples of very metal-poor stars
in Paper I observed with the Keck I HIRES (hereafter, the HIRES sample) 
and  of metal-poor dwarfs 
observed at the Okayama Astrophysical Observatory (hereafter, the OAO sample). 
We also re-analyze the  data of dwarfs observed by
 Clegg et al. (1981) and Fran{\c c}ois (1987, 1988) based on our system of
analysis to eliminate the systematic differences between our and the previous
analyses. We further perform the non-LTE (NLTE) abundance 
analysis for 
all these sample stars to examine how the  NLTE abundances of S
behave and  how the NLTE affects the LTE 
abundance determination, since a NLTE analysis of S abundance in 
metal-poor stars has not been done until now.

\section{Observations and Measurements}  

The basic data of the sample stars of HIRES, OAO, Fran{\c c}ois (1987, 1988),
and Clegg et al. (1981) are presented in Table 1. Spectral types   in the 
third column were adopted from the SIMBAD database, operated at CDS, Strasbourg, 
France. Parallaxes $\pi$ and apparent $V$ magnitudes in the fourth and
sixth columns, respectively,  were taken from the {\it Hipparcos} Catalogue 
(Perryman et al. 1997).

The CCD spectroscopic observations of the HIRES sample were carried out
in 1997 and 1999 using HIRES  on the Keck I telescope for the wavelength 
range, 6330 \AA \lsim $\lambda$ \lsim 8760 \AA, with resolution $ R = 45000$ 
and 60000,  respectively.  The journal of observations is presented in Table 1
of Paper I, and the data reduction was performed with the MAKEE package 
developed by one of us (T.A. Barlow) for HIRES data. The reader is asked 
to refer to Paper I for the details of observations and data reduction.  
The S/N ratios in the  wavelength region around the \SI lines at 8694 \AA\  
are estimated  to be 190 -- 410 in  the sample stars. The low values are  
mainly due to the difficulty of complete removal of fringes in the spectra.
The observed spectra in the vicinity of the \SI lines are shown in  Figures
 1a -- 1f.

The OAO sample of 25 dwarfs and the Moon were  observed in 1997 and 1998
using the Coude spectrograph of the 188 cm telescope at the Okayama Astrophysical
Observatory (OAO), National Astronomical Observatory of Japan. 
The wavelength range observed is $\lambda\lambda$ 8400 -- 8830 \AA, and
the resolution $ R \sim 24500 $ at 8700 \AA. The CCD data were reduced 
with the IRAF\footnote{IRAF is distributed by the National Optical
Astronomy Observatories, which is operated by the Association of
Universities for Research in Astronomy, Inc., under cooperative agreement
with the National Science Foundation.}  package, following the standard
procedure of extracting one-dimensional spectra. The S/N ratios around 
8700 \AA\  are in the range of 120 -- 400, but mostly between 200 -- 350.

Equivalent widths (\Wlam) of the two blend-free \SI lines
at 8693.958 \AA\  and 8694.641 \AA\ in spectra of the HIRES sample were 
measured by Gaussian fitting 
(or by direct integration if necessary) using the {\it splot\/} task of IRAF,
and are listed   in the HIRES sample entry of Table 5.
It turned out that 
our \Wlam values of the \SI  8694.641 \AA\ excellently agree with those 
measured in Paper I for the sample stars, except for the giant star HD 88609. 
While the \Wlam value for HD 88609 was estimated to be 1.9 m\AA\  as an
uncertain 
value in Paper I,  it should be replaced with our measurement ($< 2.2$ m\AA)
as an upper limit corresponding to the noise level with a
S/N ratio of $\sim$ 230, because the \SI 
8694.641 \AA\  line is not detected at this noise level, as seen in Figure 1c. 
Since the spectral resolution $ R$  of data of the OAO sample is not 
high enough to 
separate two \SI lines clearly, two blended \SI lines were regarded as
one line and its \Wlam was measured by a direct integration using the
{\it splot\/} task of IRAF. The results are given in the seventh column
of Table 4. We also measured \Wlam of six \FeI lines in the OAO
sample by Gaussian fitting
using the {\it splot\/} task of IRAF, and listed them in Table 3.
Values of \Wlam measured by \Francois (1987, 1988) and Clegg et al. (1981)
for two \SI lines are summarized in Table 5, and also those by 
\Francois (1987, 1988) for two or three \FeI  lines in Table 3.
The \Wlam data for \FeI  lines are not available in Clegg et al. (1981).

Our OAO sample overlaps with nine stars among the samples of
\Francois (1987, 1988) and Clegg et al. (1981), so that 
the sample of this study consists of six giants and 61 dwarfs 
(62 dwarfs, including the Sun) in total.

\section{Abundance Analyses}  

Abundance analyses were carried out using Kurucz's (1993) ATLAS9 
line-blanketed model atmospheres, based on which the atmospheric models 
of individual stars were constructed by interpolation in terms of 
atmospheric parameters of effective temperature \Teff, 
surface gravity \logg, and metallicity [Fe/H].
In this section we describe the determinations of atmospheric parameters
and the analyses of Fe and S abundances with LTE and NLTE calculations.

\subsection{Atmospheric Parameters}  

\subsubsection{Reddening Estimates}  

While interstellar reddening $E(B-V)$ affects  estimates of 
effective temperatures based on color indices, it has been usually
assumed that no reddening corrections need be applied to the stars within
100 pc. However, this  assumption could break down for some stars, 
with significant reddening (\gsim 0.01 mag), even if they are within 
100 pc, because the maps of reddening in the Galaxy show a very patchy 
distribution of reddening at all galactic latitudes (Burstein \& Heiles 
1978, 1982). 
Because $E(B-V)$=0.10 mag has been found to alter the derived \Teff 
by 400 -- 550 K for such cool stars as in our sample (Laird, Carney,
\& Latham 1988),  $E(B-V)$=0.01 may change \Teff by 40 -- 55 K.
Our test calculations confirmed these changes in \Teff.
Consequently, we decided to estimate  $E(B-V)$ for {\it all} our sample stars
regardless of their distances.

Almost all reddening corrections had been made based on the  maps of
 Burstein \& Heiles (1978, 1982) until 1998 when a new, modern source of
reddening was published by Schlegel, Finkbeiner, \& Davis (1998).
They constructed the full-sky maps of the Galactic dust using the
far-infrared data observed by the {\it Infrared Astronomy Satellite}
({\it IRAS}) and the Diffuse Infrared Background Experiment (DIRBE) 
on board the {\it COBE} satellite. They demonstrated that the new dust 
maps predict reddening with an accuracy of 16 \%, which is twice as 
accurate as that estimated from the  Burstein \& Heiles' maps in regions 
of low and moderate reddening. Hence we used the new dust maps of 
Schlegel et al. (1998) to estimate  $E(B-V)$ of our sample stars. 
After we estimated a total reddening  $E(B-V)_{\rm T}$ in the direction 
of a given star, the $E(B-V)$ to the star at distance {\it D} pc was 
calculated with the same relation as employed by Beers et al. (2000),
 $E(B-V)= E(B-V)_{\rm T}\{1 -$ exp[$-|D$ sin {\it b}$|$/{\it h}]\}, where 
{\it b} is the Galactic latitude and {\it h} is a scale height of 125 pc 
assumed for the dust layer. Distances were estimated from the {\it Hipparcos} 
parallaxes $\pi$ listed in Table 1.
Resulting  $E(B-V)$ larger than 0.01 mag are given in the fifth column of 
Table 1 and used for  determinations of \Teff and \logg.

As clearly seen from these results, reddening should be examined    
for stars even at close distances. Note a typical example of HD 190248
at a distance of $D=6.11 $ pc ($\pi=163.73$ mas) with a reddening of 
$E(B-V)=0.02$ mag in the \Francois' (1987) sample.

\subsubsection{Effective Temperatures} 

Effective temperatures of the HIRES sample, one giant star HD 111721,
and one dwarf star HD 182572 
were derived using the empirical calibration of Alonso, Arribas, \& 
Mart\'inez-Roger (1999a) for $(V-K)$ index and [Fe/H], which is based on the 
infrared flux method (IRFM).  The observed $(V-K)$ indices for these 8 stars 
were figured out by adopting $K$ magnitudes from Alonso, Arribas, 
\& Mart\'inez-Roger (1994) for HD 44007, HD 84937, HD 111721, and HD 182572, 
from Alonso, Arribas, \& Mart\'inez-Roger (1998) for HD 88609, HD 165195, and 
HD 184266, and from Carney (1983) for HD 175305. Resulting values are given 
in  Table 1.  These  $(V-K)$ indices were corrected for reddening using a 
standard relation, $E(V-K)= 2.72E(B-V)$, and applied to the empirical calibration. 
With regard to one more parameter [Fe/H] in the empirical calibration, 
we adopted the values  where  the second  decimal place
is rounded off to 5 or 0,  and listed them in the fourth column of Table 2.
The original values for  these [Fe/H] are presented in the eleventh column
of Table 1 together with the literature entered in the last column.

Effective temperatures  of the OAO, \Francois (1987, 1988), and 
Clegg et al. (1981) sample stars were estimated  using the IRFM-based 
empirical calibration (Eq. 9) of Alonso, Arribas, \& Mart\'inez-Roger (1996b) for
Str\"omgren ($b - y$) and $c_{1}$ indices and [Fe/H]. The Str\"omgren indices
were adopted from Hauck and Mermilliod (1998) and listed in the ninth and 
tenth columns of Table 1.  We used the rounded-off values of [Fe/H] the same 
as the case of HIRES sample.  Reddening corrections were applied to these 
Str\"omgren indices making use of the adopted $E(B-V)$ and the well known 
relations of $E(b-y)=0.73E(B-V)$ and $E(c_{1})=0.15E(B-V)$ (Crawford 1973).

The resulting \Teff for all our sample stars are rounded off to the second
digit (10 K) and summarized in the second column of Table 2.
 The \Teff determined by Alonso et al. (1999b) and
Alonso et al. (1996a) for giants and dwarfs, respectively, are listed in
the sixth column for comparison. The differences $\delta$\Teff $=$ 
\Teff(Ours) $-$ \Teff(Alonso et al's) are also given in the seventh column. 
Inspection of  $\delta$\Teff demonstrates that our \Teff values  
agree well with  those of Alonso et al mostly to within $\pm 100$ K, while 
discrepancies larger than 100 K are found in five stars.
This good agreement is reasonably expected since both our and the 
Alonso et al's \Teff are based on the same IRFM framework.

\subsubsection{Surface Gravities}  

Surface gravities (log $g$) were derived following the standard procedures,
 based on data of \Teff, $V$ magnitude, parallax, $E(B-V)$, 
bolometric magnitude, and the theoretical evolutionary track.

First,  we calculated the absolute visual magnitudes ($M_{\rm V}$) from 
data of  $V$ and {\it Hipparcos} parallaxes ($\pi$) adopting a $V$-band 
absorption $A_{V}=3.1E(B-V)$. We then separately estimated bolometric 
corrections of $V$, BC($V$), for giants and dwarfs. 
The BC($V$) for giants were calculated for \Teff and [Fe/H] adopted  in
 Table 2 using the Alonso et al.'s (1999a) calibrations 
(Eq. 17 or 18)  which are presented as a function of  \Teff and [Fe/H].
On the other hand, BC($V$) for dwarfs were obtained for \Teff and 
[Fe/H] adopted, interpolating the grid of BC($V$) for log $g~=4.0$ 
models calculated by Alonso, Arribas, \& Mart\'inez-Roger (1995).  
Resulting BC($V$) were applied to  estimate of absolute
bolometric magnitudes $M_{\rm bol}$ which are listed in Table 1.
 
The masses of stars were evaluated for the adopted \Teff and  $M_{\rm bol}$
 on the theoretical evolutionary tracks of the Italian group.  For the  
sample of giants and the very metal-poor dwarf HD 84937, we adopted the 
evolutionary tracks with solar-scaled mixture of abundances and  initial 
chemical compositions computed by Girardi et al. 
(1996: [$Y=0.230, Z=0.0001$] track for HD 84937, 88609, and 165195)
and Girardi et al. (2000: [$Y=0.23, Z=0.0004$] for HD 44007 and 184266;
[$Y=0.23, Z=0.001$] for HD 111721 and 175305).
In the estimates for the masses of HD 44007 and HD 165195, we failed to 
find the evolutionary tracks corresponding to the positions with \Teff and 
$M_{\rm bol}$ on the HR diagram, so we assumed 0.6 $M_{\odot}$ for 
these two stars.  This assumption is probably adequate since a mass range 
indicated from masses obtained for our sample of remaining giants is  
0.6 -- 1.0 $M_{\odot}$. 

For the sample of remaining dwarfs,  the masses were inferred on the 
theoretical evolutionary tracks computed by Salasnich et al. (2000).
We used their evolutionary tracks with $\alpha$-enhanced mixture of 
abundances and  initial chemical compositions of [$Y=0.250, Z=0.008$], 
[$Y=0.273, Z=0.019$], and [$Y=0.320, Z=0.04$] for our sample stars
with [Fe/H] $< -0.1$, $-0.1 \leq$ [Fe/H] $\leq +0.1$, and [Fe/H] $> +0.1$,
respectively.

In these procedures of \logg estimate, we adopted the following
values for the Sun: $T_{\rm eff},_{\odot} = 5780$ K, \logg$_{\odot} = 4.44$,
and $M_{\rm bol},_{\odot} = 4.74$. The log $g$ values thus derived are 
summarized in the third column of Table 2.

\subsubsection{Microturbulences}  

Microturbulences ($\xi$) of the stars in the HIRES sample were determined 
based on \Wlam data of \FeI lines measured  in Paper I, eliminating 
any trend of abundances with line strength.  The values of $\xi$ for all 
samples of dwarfs were calculated using the empirical relation,

\vskip 0.5cm

$\xi$ (\kms)$ = 1.25 + 8 \times 10^{-4}$(\Teff$ - 6000) - 1.3($\logg$ - 4.5)$,

\vskip 0.5cm
which was found by Edvardsson et al. (1993).
As for the $\xi$ of the giant star HD 111721, it was taken from Ryan \&
Lambert (1995) because this relation cannot be applied for giants.
Resulting values of $\xi$ are listed in the fifth column of Table 2.

\subsection{LTE Analyses}  

We used the WIDTH9 program written by R. L. Kurucz to determine the LTE  
abundances of S and Fe based on the adopted model atmospheres.

\subsubsection{Fe Abundances}  

The Fe abundances of the HIRES sample were derived from both \FeI and \FeII
lines by analyzing their \Wlam given in Paper I, while those of the OAO and 
\Francois (1987, 1988) samples were obtained  from the measured \Wlam of 
selected \FeI lines.  The {\it gf}\ values compiled by Kurucz (1995) were 
used. The enhancement factor, $f_{6}$, which should be multiplied by the 
classical van der Waals damping constant, was estimated to be 1.2 for \FeI 
lines with lower excitation potentials ($\chi$) less than 2.6 eV using 
the empirical calibration by Simmons \& Blackwell (1982), and 
$f_{6} = 1.4$ was adopted from Edvardsson et al. (1993) for lines with 
$\chi > 2.6$ eV.

The results of Fe abundances yielded from the HIRES sample are presented 
in Table 6 as  [Fe~{\sc i}/H]$_{\rm LTE}$ and [Fe~{\sc ii}/H]$_{\rm LTE}$,
which were calculated relative to the solar value of log~Fe$_{\odot} = 
7.51$ (Holweger, Kock, \& Bard 1995). Hereafter, we will separately deal 
with the abundances obtained from \FeI and \FeII lines for the HIRES sample.
 
Resulting Fe abundances of the OAO and \Francois (1987, 1988) samples 
are summarized in Table 3 together with \Wlam, the {\it gf} values,  and 
the lower excitation potentials ($\chi$). The averages of log~Fe are also 
listed as log~Fe {\sc i}$_{\rm LTE}$ in Table 7. The values of 
[Fe~{\sc i}/H]$_{\rm LTE}$ of the OAO sample were calculated relative to
our solar value of log~Fe$_{\odot} = 7.41$ to eliminate the systematic
errors due to uncertainties of the log~$gf$ and $f_{6}$ values, while
those of the \Francois (1987, 1988) sample relative to log~Fe$_{\odot} = 
 7.51$. Those of Clegg et al. (1981) were simply adopted from the 
values analyzed by them and are listed in Table 7. 
The [Fe~{\sc i}/H]$_{\rm LTE}$  of the giant star HD 111721 is included 
in Table 6.

\subsubsection{S Abundances}  

The S abundances (log S) were calculated adopting the $gf$ values compiled
by Kurucz (1995): log~$gf = +0.080$ for the 8694.641 \AA\  line and 
log~$gf = -0.510$ for the 8693.958 \AA\  line.  The lower excitation 
potential for both lines is $\chi = 7.87$ eV, which means that the \SI 
lines analyzed are high excitation lines and so the abundances
derived from them are sensitive to errors of \Teff rather than \logg 
especially when the lines are considerably weaker (\lsim 20 m\AA)
and \Teff is cooler than about 5500 K. The enhancement factor $f_{6} = 
2.5$ was adopted from Feltzing \& Gonzalez (2001).

The log~S of the OAO sample stars were derived from \Wlam of the blended \SI
line feature which is regarded as one line, and are shown in the last
column of Table 4 and also in  Table 7 as log~S$_{\rm LTE}$.
The S abundances of the HIRES sample and the samples of \Francois 
(1987, 1988) and Clegg et al. (1981) were obtained from the two \SI lines, 
and are listed in Table 5.  The averaged abundances for the samples of 
\Francois (1987, 1988) and Clegg et al. (1981) are also given in Table 7 
as log~S$_{\rm LTE}$.

The values of [S/H]$_{\rm LTE}$ for the HIRES sample and the samples of
 \Francois (1987, 1988) and Clegg et al. (1981) were calculated relative 
to the solar value of log~S$_{\odot} = 7.21 $ (Anders \& Grevesse 1989), 
while those of the OAO sample were calculated relative to our solar 
value of log~S$_{\odot} = 7.22$.
These results are listed in Tables 6 and 7 for the HIRES sample and 
the giant star HD 111721 and for all the remaining dwarfs, respectively.

The synthetic line profiles of \SI lines at 8693.9 \AA\  and 8694.6 \AA\ 
were computed for the HIRES sample stars using the adopted S abundances,
and overplotted in Figures 1a -- 1f. Except for the upper limit case
of HD 88609, the synthetic profiles of \SI 8694.6 \AA\ line for the 
remaining stars fit well with the  observed ones, while the observed 
profile of HD 84937 (Figure 1b) slightly disagrees with the synthetic one 
on the blue side of the profile which seems  asymmetric relative to the 
synthetic one.  The line profile of this star may be influenced by a 
rather low S/N ratio ($\sim 230$) and incomplete removal of the fringe 
pattern.  Although we confidently regard the absorption feature 
at 8694.6 \AA\ observed in HD 84937 as the \SI line, we should confirm the 
line with a follow-up observation.

\subsubsection{[S/Fe]}  

The values of [S/Fe] for the HIRES sample were computed from the 
above-obtained [Fe~{\sc i}/H]$_{\rm LTE}$, [Fe~{\sc ii}/H]$_{\rm LTE}$, 
and [S/H]$_{\rm LTE}$, and are given as [S/Fe~{\sc i}]$_{\rm LTE}$ and 
[S/Fe~{\sc ii}]$_{\rm LTE}$ in the fifth and sixth columns of Table 6, 
respectively, together with the [S/Fe~{\sc i}]$_{\rm LTE}$ of the giant 
star HD 111721.  The error bars are discussed in \S 3.4.

In Table 7, the [S/Fe~{\sc i}]$_{\rm LTE}$ results are listed in 
the sixth column for the dwarf samples of OAO, \Francois (1987, 1988), 
and Clegg et al. (1981).

\subsection{NLTE Analyses}   

The NLTE abundances of S were computed for the two \SI lines following the
same procedures as described in Takada-Hidai \& Takeda (1996).
The grids of NLTE corrections were constructed for the parameter ranges of
\Teff $=4500$ -- $6500$ K, \logg$ = 1.0$ -- $5.0$, and 
[Fe/H]$ = 0.0$ -- $-3.0$, assuming a constant microturbulence 
of 2.0 \kms\  and changing enhancement  of S abundances. 
The grids for  \SI 8693.9 and \SI 8694.6 lines are presented in Tables A1
and A2 in the Appendix, respectively.  The NLTE corrections for the S 
abundance defined as 
$\Delta({\rm S}) \equiv$ log~S$_{\rm NLTE}$ $-$ log~S$_{\rm LTE}$  
were evaluated on these grids for two \SI lines of the HIRES sample and 
the samples of \Francois (1987, 1988) and Clegg et al. (1981) using their 
measured \Wlam.  As for the OAO sample, since the \SI lines are blended, 
we computed the \Wlam of each \SI line corresponding to the derived 
log~S$_{\rm LTE}$ and used them to estimate $\Delta({\rm S})$. 
The average values of each $\Delta({\rm S})$ are given in the seventh column
of Tables 6 and 7 for the HIRES sample and the giant star HD 111721
and for the sample of dwarfs, respectively.

The NLTE abundances of Fe were calculated for [Fe {\sc i}/H]$_{\rm LTE}$
in all our samples using the polynomial relation derived by 
Israelian et al. (2001), which is based on the NLTE work of 
Th\'evenin \& Idiart (1999).  The results are shown as 
[Fe {\sc i}/H]$_{\rm NLTE}$  in  Tables  6 and 7. 

The values of [S/Fe {\sc i}]$_{\rm NLTE}$ were computed 
based on the above-obtained [S/H]$_{\rm LTE}$,  $\Delta({\rm S})$, 
and [Fe {\sc i}/H]$_{\rm NLTE}$, and are summarized in Tables 6 and 7
for each sample.
 
Since Fe abundances inferred from \FeII lines are separately dealt with
in the HIRES sample, the results of [S/Fe {\sc ii}]$_{\rm NLTE}$ 
are also entered in Table 6. Here we applied the values of
[Fe {\sc ii}/H]$_{\rm LTE}$ to those of [S/H]$_{\rm NLTE}$ to get these 
results because an LTE Fe abundance deduced from \FeII has been found 
to be free from NLTE effects and reliable enough as suggested, for example,
by Lambert et al. (1996) in their abundance study of RR Lyrae stars
 and by Th\'evenin \& Idiart (1999) and Gratton
et al. (1999) in studies of the NLTE effect on \FeII in metal-poor stars.

\subsection{Error Analyses}  

Although there are many factors which yield errors in the abundances, 
we focus only on the main factors of uncertainties of \Teff, \logg, and $\xi$.

The uncertainties of \Teff for  all our samples were estimated 
to be $\Delta$\Teff$= \pm 100$ K, taking into account the \Teff differences 
($\delta$\Teff) shown in Table 2, since most of our \Teff values agree well
with those of Alonso et al. (1996a, 1999b) within $\pm 100$ K.

The uncertainties of \logg  are mainly caused by  errors in \Teff,
stellar mass, and  $M_{\rm bol}$.
Test calculations have found that  $\Delta$\Teff$=  100$ K corresponds to 
$\Delta$\logg $\sim 0.06$ dex.  Errors in the mass are introduced from a 
selection of theoretical evolutionary tracks for fixed \Teff and  
$M_{\rm bol}$ (i.e. luminosity). Mass errors of $\sim 0.05$M$_{\odot}$ 
estimated on the evolutionary tracks corresponds to  $\Delta$\logg $\sim 
0.05$ dex.  Errors of $M_{\rm bol}$ essentially come from the parallax errors.
The typical error of 3 \% for the parallax yields $\Delta$\logg $\sim 
0.06$ dex.  The quadratic sum of these uncertainties in \logg amounts to 
$\pm 0.1$ dex.  Allowing 0.05 dex for other possible error sources in 
the estimation procedures, we simply added this value to the quadratic sum
and adopted  $\Delta$\logg $= \pm 0.15$ dex as the total uncertainty.

The uncertainties of  microturbulences $\xi$  for the HIRES sample were 
deduced from test calculations to examine whether \FeI abundances show 
any trend with line strength for different values of $\xi$.
Since  $\xi$ for the samples of OAO, \Francois, and Clegg et al. 
were estimated using the empirical relation of Edvardsson et al. (1993), 
the uncertainty of $\Delta\xi \simeq 0.22$ \kms\ was estimated for the 
above uncertainties of \Teff and \logg. The rms scatter for this relation 
was suggested to be about 0.3 \kms\  by Edvardsson et al. (1993), so that 
the total uncertainty became $\pm 0.37$ \kms.  We then adopted   
$\Delta\xi = \pm 0.5$ \kms\ as the total uncertainty allowing for other 
possible errors.

The abundance errors of S and Fe for the HIRES sample and HD 111721  were  
calculated for these uncertainties, and are given in Table 6 as the error 
bars on the [S/Fe] values. The abundance errors for the remaining dwarfs 
sample were also evaluated and listed in Table 8.  We regard the combined 
error of $\pm 0.16$ dex as a typical error for [S/Fe {\sc i}] values in 
the sample of dwarfs.

\section{Results and Discussion}  

The results of abundance analyses of S and Fe are summarized in Tables 6 
and 7 for six stars of the HIRES sample and the giant star HD 111721 and 
for 61 dwarfs, including the Sun, respectively. 
Since nine stars of the OAO sample overlap with those of the samples of 
\Francois (1987, 1988; four stars) and Clegg et al. (1981; five stars), 
the abundance results of the OAO sample were preferentially adopted for 
these  stars.  We will describe these results below and discuss
the results for S and Fe abundances of all our samples. 

\subsection{NLTE Corrections}   

While the NLTE corrections of the S abundances, $\Delta$(S), are found 
to be in the range of $-0.09$ -- 0.00 dex for all of our sample,  
most of them concentrate on the range of  $-0.01$ -- $-0.03$ dex. 
Consequently, neglect of NLTE effect does not produce significant errors 
leading to the wrong conclusions of behavior of S.
On the other hand, as seen from the comparison of [Fe {\sc i}/H]$_{\rm LTE}$ 
with [Fe {\sc i}/H]$_{\rm NLTE}$, NLTE corrections of \FeI abundances are
found to be considerably larger with the  range of $-0.09$ -- $+0.29$ dex.
The negative correction values in the range of $-0.09$ -- $-0.02$ dex are 
assigned only to the nine metal-rich stars, but all positive ones are
for our sample of metal-poor stars, which distribute mostly in the range
of $+0.05$ -- $+0.15$ dex and $+0.20$ -- $+0.29$ dex among the samples of
dwarfs and giants, respectively.
These NLTE corrections for \FeI abundance yield significant changes in 
[S/Fe {\sc i}], so that they should be examined  when we investigate the  
behavior of [S/Fe {\sc i}] against  [Fe {\sc i}/H].  However, we should 
note that Gratton et al. (1999) computed NLTE corrections of \FeI abundances 
for dwarfs (\logg = 4.5) and low gravity (\logg = 1.5 and 2.25) stars in 
the metallicity range of $-3$ -- $0$ dex, and concluded that NLTE 
corrections are very small (mostly $< 0.05$ dex) in dwarfs of \Teff less 
than 7000 K, while those in low gravity stars are less than 0.4 dex for 
\Teff $< 6000$ K and the metallicity range of $-1$ -- $-3$ dex. 
Using their Figure 9, NLTE corrections for our giants sample were estimated 
to be less than about 0.15 dex, which are systematically smaller than those 
adopted in this study. Since  [Fe/H] values of cool metal-poor stars are 
mainly determined from \FeI lines, further examinations of NLTE effect on 
\FeI abundance are worth performing. 

As mentioned in section \S 3.3, the abundances derived from \FeII lines
can be regarded as being reliably free from NLTE effects.  Hence analyses 
of \FeII lines are recommended whenever such data are available.

\subsection{Behavior of  [S/Fe]}  

To clarify the behavior of S in the metallicity range $- 3$ \lsim
[Fe/H] \lsim $+0.5$, we plotted  [S/Fe]$_{\rm LTE}$ and [S/Fe]$_{\rm NLTE}$
 against  [Fe/H]$_{\rm LTE}$ and [Fe/H]$_{\rm NLTE}$  in Figures 2 and 3, 
respectively. 

We first deal with the LTE behavior of [S/Fe] shown  in Figure 2.
As for [S/Fe {\sc i}] in all our samples of dwarfs and giants, 
it shows an increasing trend as [Fe {\sc i}/H] decreases. 
A slope of this trend is calculated to be $-0.27 \pm0.15$ 
(where $\pm0.15$ are 1 $\sigma$ errors, and these errors are
also given to the slopes yielded from  a least-square linear fit in other
cases) 
by a least-square linear fit for all the [S/Fe {\sc i}] results derived 
from the  atmospheric models  adopted in this study, except for an upper 
limit of HD 88609. The upper limit results of HD 88609 are not considered 
for the least-square linear fits in both cases of LTE and NLTE.
This linear fit gives [S/Fe {\sc i}] $\sim 0.7$ at [Fe {\sc i}/H]$=-2.5$,
while [S/Fe {\sc i}]$\sim 1.2 $ dex of HD 165195 deviates from the fit by 
about 0.5 dex around the same metallicity. The high value of   HD 165195
seems to occur from the adopted \Teff of 4190 K which is on the lowest
boundary of a range of \Teff (4131 -- 4507 K) previously determined 
(see Paper I).  This low \Teff produces the high S abundance derived from 
high-excitation lines and the lower \FeI abundance which breaks ionization 
equilibrium between \FeI and \FeII by a factor of 0.5 dex.  
Our test calculations showed that \logg\ should be lowered by about 0.5 
to obtain ionization equilibrium between \FeI and \FeII with abundance 
differences of 0.3 dex.   However, such low value of \logg $\sim 0.5$ may 
not be valid since it is inferred from the mass of $\sim 0.2 M_{\odot}$, 
which seems unreasonable for HD 165195, when \Teff and  $M_{\rm bol}$ are 
fixed.  If we calculate the  [S/Fe {\sc i}] value based on the atmospheric 
model determined by Pilachowsky et al. (1996) and adopted in Paper I, 
this results in $+0.58 \pm 0.28$ dex as listed in the Pap.I entry of Table 6,
and plotted with a filled asterisk in Figure 2.  Adopting this Pap.I result 
of [S/Fe {\sc i}], a least-square linear fit yields a slope of $-0.23 
\pm0.13$, which is almost the same as the above-obtained slope of $-0.27$.
 
The  [S/Fe {\sc ii}] values  of the HIRES sample are also plotted in Figure 2
with half-filled diamond.  In a case of HD 165195, the value listed in 
the  Pap.I entry of Table 6 is also plotted with an open asterisk, which 
is located very close to the point inferred from this study (the Ours entry 
of Table 6).   Combining these HIRES data with the [S/Fe {\sc i}] data of 
OAO, \Francois (1987, 1988), and Clegg et al. (1981), it is found that 
[S/Fe] shows a continuous increase  with a slope of   $-0.23 \pm0.13$ as 
[Fe/H] decreases.  This trend supports the  one found in the case of 
[S/Fe {\sc i}].

Now, we inspect the NLTE  behavior of [S/Fe] depicted in Figure 3.
Because the NLTE anaysis seems to be more reliable than the LTE one, 
we preferentially adopt the NLTE results as our final results in this study.
As concerns [S/Fe {\sc i}] of all our dwarfs and giants samples, it shows 
the same trend as in the above LTE case.  However, since significant NLTE 
corrections to \FeI abundances make \FeI abundances of metal-poor stars 
higher and [S/Fe {\sc i}] lower, the slope becomes  $-0.17 \pm0.15$, which 
is flatter than the LTE case, as illustrated by the least-square linear fit 
drawn in Figure 3 with dashed line.
If we consider the [S/Fe {\sc ii}] values  of the HIRES sample together
with [S/Fe {\sc i}] data of OAO, \Francois (1987, 1988), and Clegg et al. 
(1981), we find a slope of $-0.19 \pm0.14$. 
Both slopes remain almost unchanged for the [S/Fe] points in the  Pap.I entry
of Table 6 for the case of HD 165195.

Judging from the trends of LTE and NLTE behaviors of [S/Fe]
against [Fe/H] observed in all our sample stars, we may safely conclude
that  [S/Fe] increases progressively and continuously with a slope of 
$\sim -0.2$ as [Fe/H] decreases from $+0.5$ dex to $-3$ dex, though
the observed data are distributed with a range of scatter  of
$0.3$ -- $0.5$ dex. Our conclusion is qualitatively consistent with that 
of Israelian \& Rebolo (2001). They found that [S/Fe] shows an increase
trend with the slope of $-0.46 \pm0.06$ which is roughly twice as 
steep as those found in our sample stars.
If we combine the results of six stars observed by  Israelian \& Rebolo 
(2001) with those of our samples and calculate the slope of increase trend 
of [S/Fe {\sc i}], we obtain the slope of $-0.25 \pm0.17$, 
which is significant at the 1.5 $\sigma$ level.
This least-square linear fit is depicted with solid line in Figure 3, 
together with Israelian \& Rebolo's (2001) data plotted with a double circle. 
The same slope of $-0.25 \pm0.15$, which is significant at the 1.7  $\sigma$
level,  is also derived from the combination of 
data of Israelian \& Rebolo (2001) and our data which include the results of 
[S/Fe {\sc ii}] instead of [S/Fe {\sc i}] in the HIRES sample.  
Although this slope, $-0.25$,  of the trend is significant only
at the 1.5 -- 1.7 $\sigma$ level and
still flatter than that of  Israelian \& Rebolo (2001), 
it is more favorable to our above conclusion. 
As for a steeper slope found by Israelian \& Rebolo (2001), which is 
significant at the 7.7 $\sigma$ level, it might be influenced by a bias in
 a smaller number of their sample (26 stars in their Figure 3) in 
comparison with our sample of 67 stars [73 stars in our Figure 3, including 
6 stars of Israelian \& Rebolo (2001)].  A slope is essentially determined by
[S/Fe]  in halo stars, so that further observations of a larger sample of 
halo stars are indispensable to establish an increase trend of [S/Fe] with 
a trustworthy value of slope.
It is also interesting that the slope of $-0.25$ is comparable to  
those ($\sim -0.3$)  found for [O/Fe] (e.g. Paper I; Israelian et al. 2001).
As discussed below, since O and S 
are volatile elements, it may be plausible to expect that they show a similar 
behavior against metallicity, [Fe/H].

On the contrary, the linearly increasing trend of [S/Fe] is not consistent 
with the conclusion suggested by \Francois (1988) that [S/Fe] forms a 
plateau in halo stars.  His conclusion seems to be affected by a bias in 
that his sample does not contain halo stars with  [Fe/H] \lsim $-1.5$.

Now, we briefly discuss our results in relation with the theoretical 
studies of chemical evolution of the Galaxy.  
Theoretical predictions based on standard SNe II and Ia models  may not explain
our linearly increasing trend of [S/Fe] in the range of [Fe/H] \lsim $-1$
(e.g. Chiappini et al. 1999; Goswami \& Prantzos 2000), however, it 
may be possible to explain the  observed trend of [S/Fe] if we consider
the explosive nucleosynthesis in ``hypernovae'', i.e., SNe with
very large explosion energies of $E=$(10 -- 100)$\times 10^{51}$ ergs,
proposed by Nomoto et al. (2001) and studied in detail by 
Nakamura et al. (2001).  Nakamura et al. (2001) carried out detailed 
nucleosynthesis calculations for hypernovae with these energies as well 
as for ordinary core-collapse SNe with $E=1 \times 10^{51}$ ergs  
for comparison. They found that a larger amount of S is synthesized 
by oxygen burning in hypernovae, which leads to higher [S/Fe] ratios to 
be observed in metal-poor halo stars if hypernovae occurred in the early 
phase of the Galaxy evolution.
For example, inspection of  Table 2 or 3 of Nakamura et al. (2001)
suggests that the ratios of [S/Fe] \gsim 1 may be attained in the
ejecta of hypernovae with $E$ \gsim $10^{52}$ ergs in the metallicity
range of [Fe/H] $< -1$. They also found that even one hypernova can 
produce 2--10 times more Fe than normal core-collapse SNe, which 
makes [$\alpha$-elements/Fe] ratios smaller. Consequently, if such metal-poor
halo stars, as included in our sample, formed from some hypernova ejecta, 
the iron mass of these stars should be smaller by factors of \gsim 10 than 
those presented, for instance, in Table 2 to explain the [S/Fe] trend 
observed  in this study and Israelian \& Rebolo (2001). 
As discussed for the case of abundances in the black hole binary GRO 
J1655$-$40 by Nakamura et al. (2001), there may be some possible ways to 
produce the ejecta with smaller Fe mass using hypernova models with a 
certain mass cut at large $M_r$ (mass included in the radius $r$ of the 
precollapse star) or with asymmetric explosions such as have a jet. In 
addition to these possibilities, the mixing and dilution of ejecta to the 
interstellar medium plays an important role in determining the metallicity 
of halo stars.  Nakamura et al. (1999) suggested that the [Fe/H] of halo 
stars are mainly determined by the mass of interstellar hydrogen mixed with 
the ejecta of the relevant SN II, and that the mass of such interstellar 
hydrogen varies by an order of magnitude and is larger for both cases
of the larger explosion energy of SN and the larger Str\"omgren radius of the 
progenitor.  A more detailed discussion on the justification of the 
increasing trend of [S/Fe] has been made with relation to 
hypernovae/supernovae by Israelian \& Rebolo (2001).

Another possibile explanation for the increasing trend of [S/Fe] at low [Fe/H] 
has been proposed by Ramaty et al. (2000) and Ramaty,  Lingenfelter,
 \& Kozlovsky (2000) in connection with their explanation of a similar
increasing trend of [O/Fe] (cf.  Israelian et al. 2001). 
Taking into account the delayed deposition of the SN products
into the interstellar medium due to differences in transport and mixing,
which are inferred from the different characteristics of volatile (O) and
refractory (Fe) elements with dust grains,
they simulated the evolution of [O/Fe] versus [Fe/H] for the case of
a short mixing delay time (1 Myr) for O and a longer one (30 Myr)
for Fe. They found that [O/Fe] should increase monotonically up to
$\sim 1$ at [Fe/H] $\sim -3$ with a slope consistent with the previously 
obeserved one (eg. Israelian et al. 2001), and then predicted that the 
similar trend should be observed for another volatile element of S, which 
is just the demonstrated case in this study.

\section{Conclusions}  

LTE and NLTE abundances of sulfur in 6 metal-poor giants and 61 dwarfs 
(62 dwarfs, including the Sun) were explored in the range of 
$-3$ \lsim [Fe/H] \lsim $+0.5$ using \SI 8693.9 \AA\  and 8694.6 \AA\  lines. 
NLTE effects in the S abundances are found to be small and practically 
negligible.
The behavior of [S/Fe] vs. [Fe/H] exhibits a linearly increasing trend
without plateau with decreasing [Fe/H]. Although the slope of the linearly
increasing trend is found to be in the range of $-0.17$ -- $-0.25$ 
for the NLTE behavior, the value of $-0.25$ is the most favorable one
for all observed data used in this study. It is interesting to note that 
this slope is comparable to that ($\sim -0.3$) observed in [O/Fe],
which may be plausible since S and O are both volatile elements.

The observed behavior of S may require chemical evolution models of the 
Galaxy, in which scenarios of hypernovae  nucleosynthesis and/or 
time-delayed deposition into interstellar medium are incorporated.

Since our conclusions are essentially based on the small sample of halo stars,
further observations should  be performed for a larger sample of halo stars
 to establish our conclusions and explore 
the behavior of S in the very beginning stage of the chemical evolution
of the Galaxy.



\acknowledgments

We would like to thank  G. Israelian for his helpful comments and discussions,
and the referees, S. G. Ryan and P. \Francois, for their comments
which helped us to improve the paper. We also wish to thank D. J. Schlegel
and T. C. Beers for their kind help with handling of the dust maps, 
J. X. Prochaska and M. Asplund for the comments,
and K. Osada for his great help with revision of atmospheric parameters.

We are grateful to the staff of the Okayama Astrophysical Observatory
and the W. M. Keck Observatory for their help with observations.
One of us (MTH) acknowledges the financial supports from grant-in-aid
for the scientific research (A-2, No. 10044103) by Japan Society
for the Promotion of Science as well as from Tokai University in 1999
fiscal year, which enabled his observation with HIRES. 

This work is  partially supported from grant-in-aid for the scientific 
research by Japan Society for the Promotion of Science for MTH 
(C-2, No. 13640246).

This research has made use of the SIMBAD database, operated at CDS, 
Strasbourg, France.

\vskip 1cm

\appendix

\section{Appendix}

The results of NLTE calculations for two \SI lines at 8693.9 \AA\  and 
8694.6 \AA\  are given in Tables A1 and A2, respectively. 

The meanings of each column are as follows:

\begin{enumerate}
\item The 1st column: Code  stands for models with coded atmospheric 
parameters of \Teff, \logg, and metallicity. For example, {\it t65g50m0}
stands for a model with \Teff$=6500$ K, \logg$=5.0$, and [Fe/H]$=0.0$;
{\it t50g40m2} a model with \Teff$=5000$ K,  \logg$=4.0$, and  [Fe/H]$=-2.0$.

\item The 2nd column: $\lambda$ is the wavelength of the \SI line.

\item The 3rd column: $\xi$  is the microturbulence.

\item The 4th column: [S/Fe]$_{\rm i}$ stands for the input value of [S/Fe]
corresponding to $A_{\rm input}$ of S. The solar value of Fe adopted is 7.51.

\item The 5th column:  $A_{\rm input}$ is the input value of S abundance
for theoretical calculations of equivalent widths with LTE and NLTE.

\item The 6th column: $W$(LTE) is the theoretical  LTE equivalent widths
 calculated with $A_{\rm input}$.

\item The 7th column: $W$(NLTE) is the theoretical  NLTE equivalent widths
 calculated with $A_{\rm input}$.

\item The 8th column: $A$(NLTE) is the abundance of S calculated from $W$(NLTE)
under the assumption of NLTE.
Although $A$(NLTE) should be equal to  $A_{\rm input}$ in the strict sense,
there are practically small (and negligible) discrepancy due to  
numerical problems in computation. 

\item The 9th column:  $A$(LTE) is the abundance of S calculated from $W$(NLTE)
under the assumption of LTE.

\item The 10th column: $\Delta$ is the NLTE correction defined as $\Delta \equiv$
 $A$(NLTE) $-$ $A$(LTE). 

\end{enumerate}





\clearpage



\begin{center}
{\bf Figure Captions}
\end{center}

Fig. 1a. --- Observed and synthetic spectra in the vicinity
of two \SI lines for HD 44007, where observed data are shown by filled 
circles, and synthetic spectrum computed using the adopted LTE S abundance is
overplotted with solid line.

Fig. 1b. --- The same as Fig.1a, but for HD 84937.

Fig. 1c. --- The same as Fig.1a, but for HD 88609.

Fig. 1d. --- The same as Fig.1a, but for HD 165195.

Fig. 1e. --- The same as Fig.1a, but for HD 175305

Fig. 1f. --- The same as Fig.1a, but for HD 184266.

Fig. 2.  --- Behavior of sulfur with respect to iron in LTE results.
    The results of [S/Fe {\sc i}] and [S/Fe {\sc ii}] calculated for HD 165195
using the model atmosphere adopted in Paper I are plotted with filled 
and open asterisks, respectively.

Fig. 3.  ---  Behavior of sulfur with respect to iron in the NLTE results.
The results of [S/Fe {\sc i}] and [S/Fe {\sc ii}] calculated for HD 165195
using the model atmosphere adopted in Paper I are plotted the same as Figure 2.
The least-square linear fit with a slope of $-0.17$
obtained for [S/Fe {\sc i}] results of all our samples of dwarfs and giants is
illustrated by dashed line, while the same fit with a slope of $-0.25$ is
shown by solid line, which is
derived from  [S/Fe {\sc i}] data  of all our samples together  with those
of Israelian \& Rebolo (2001) plotted with double circle.


\end{document}